\documentclass{moriond}

\usepackage{booktabs}
\usepackage[separate-uncertainty = true, multi-part-units = single]{siunitx}
\usepackage{diagbox}
\usepackage{tabularx}
\usepackage{comment}
\usepackage{multirow}

\def\ra{\rightarrow}

\def\be{\begin{equation}}
\def\ee{\end{equation}}
\def\bea{\begin{eqnarray}}
\def\eea{\end{eqnarray}}

\input{belle2-symbols}
\begin{document}

\vspace*{4cm}
\title{$\mathcal{B}(B\ra \tau\nu)$ measurement with the hadronic FEI at Belle~II}

\author{Giovanni Gaudino on behalf of Belle II collaboration}

\address{Scuola Superiore Meridionale\\ 
INFN Sezione di Napoli}

\maketitle\abstracts{
We presented at Moriond EW 2025 a measurement of the branching fraction of \Btaunu decays using $\SI{387\pm6e6}{}$ \FourS collected between 2019 and 2022 with the Belle~II detector at the SuperKEKB \epem collider. We reconstruct the accompanying \Bm meson using the hadronic tagging method, while \Btaunu candidates are identified in the recoil. We find evidence for \Btaunu decays at 3.0 standard deviations, including systematic uncertainties. The measured branching fraction is $\brbtaunu = [1.24 \pm 0.41 \stat \pm 0.19 \syst] \times 10^{-4}$. The full analysis is explained in Ref.~\cite{Belle-II:2025ruy}}

\section{Introduction}

The purely leptonic decay \Btaunu\footnote{Charge-conjugate decays are implied throughout.} offers a theoretically clean probe of the Standard Model (SM), with a branching fraction given by
\begin{equation}
    \mathcal{B}(\Btaunu) = \frac{G_F^2 m_B m_\tau^2}{8\pi} \left(1-\frac{m_\tau^2}{m_B^2}\right)^2 f_B^2 |V_{ub}|^2 \tau_B,
\end{equation}
where all parameters are experimentally measured~\cite{2024pdg}, except for the decay constant $f_B$, determined via lattice QCD~\cite{aoki2024flagreview2024}.\\
Assuming SM predictions, this decay enables a clean determination of $|V_{ub}|$, independent of semileptonic $B\to X_u\ell\nu$ decays~\cite{banerjee2024averagesbhadronchadrontaulepton}. Theoretical uncertainties are currently below 1\%~\cite{aoki2024flagreview2024}. Furthermore, \Btaunu is sensitive to new physics, such as contributions from two-Higgs-doublet models or supersymmetric extensions, which can alter the branching ratio by factors up to four~\cite{Jung_2010,Crivellin:2012ye,Haller:2018nnx}.\\
The current world average is $(1.09 \pm 0.24)\times 10^{-4}$~\cite{2024pdg}. This analysis uses \SI{365.4\pm1.7}{\invfb} of data collected by \belletwo between 2019 and 2022 at the $\FourS$ resonance~\cite{Belle-II:2024vuc}, corresponding to $(387 \pm 6) \times 10^6$ \BBbar pairs, and $42.3~\invfb$ of off-resonance data to constrain continuum and $\tau^+\tau^-$ backgrounds. The \Bm meson is fully reconstructed in hadronic decays (\btag), and signal candidates (\bsig) are sought in four $\tau^+$ decay channels: $e^+\nu_e\bar{\nu}_\tau$, $\mu^+\nu_\mu\bar{\nu}_\tau$, $\pi^+\bar{\nu}_\tau$, and $\rho^+\bar{\nu}_\tau$, which together account for about 72\% of all $\tau$ decays~\cite{2024pdg}.\\
Signal selection is optimized using simulation, validated with control samples, and the branching fraction is extracted via a simultaneous two-dimensional fit to the residual calorimeter energy and missing mass squared distributions.

\section{Calibration and Model Validation}
\label{sec:calib}

In this section, we describe the efficiency corrections applied to the signal and \BBbar simulations, as well as the calibration and validation of the \eextra observable using several control samples and signal sidebands.

\subsection{Efficiency and Multiplicity Corrections}

The reconstruction efficiency of the hadronic FEI algorithm is initially estimated from simulation and corrected using two data-driven control samples: inclusive semileptonic decays (\(\Bp \to X \ell^+ \nu_\ell\)) and hadronic decays (\(\Bp \to \overline{D}^{0(*)} \pi^+\)). Correction factors, ranging from 0.6 to 1.1, are derived by comparing yields in data and simulation, and depend on the \btag decay mode.\\
Additional corrections account for differences in particle identification between data and simulation. These are obtained from control samples including leptons from \jpsi decays and low-multiplicity processes, as well as pions from \(\KS\), \(\Dstarp\), and \(\Lambda\) decays. Corrections depend on the momentum and polar angle of the track.\\
Simulation accurately models the \eextra shape within each \ng bin but underestimates or overestimates the overall multiplicity distribution. A bin-by-bin reweighting is applied to correct the \ng spectrum using different control samples.
Correction factors range from 0.8 to 1.2. The reweighting is validated in several control regions, including high \eextra, low $M_\text{bc}$, low \missM (leptonic channels), and low $P_{\text cand}$ (hadronic channels) sidebands.

\section{Signal Extraction}
\label{sec:sig_sel_opt}

To discriminate signal from background, we exploit the \eextra and \missM variables. Signal events are characterized by low \eextra and large \missM values, whereas background events typically exhibit higher \eextra and lower \missM. To capture these behaviors and their correlation, we construct a two-dimensional binned probability density function (PDF) of \eextra and \missM.\\
The fit includes five floating parameters: the common branching fraction \brbtaunu and the total background yield $n_{b,k}$ for each $\tau^+$ decay mode $k=\ep, \mup, \pip, \rhop$. The signal yields $n_{s,k}$ are not free parameters, but are derived from \brbtaunu and fixed quantities as:
\begin{equation}
n_{s,k} = 2n_{\BpBm}\times \epsilon_k \times \brbtaunu,
\end{equation}
where $n_{\BpBm} = \nBB f^{+-}$, and $f^{+-} = 0.5113 ^{+0.0073}_{-0.0108}$ is the branching fraction $\mathcal{B}(\FourS\to\BpBm)$, taken from Ref.~\cite{banerjee2024averagesbhadronchadrontaulepton}. The efficiency $\epsilon_k$ is the probability to reconstruct a \Btaunu event in category $k$, averaged over all true $\tau^+$ decay modes. It includes the $\tau^+$ branching fractions and reconstruction cross-feed, as predicted by simulation.\\
We perform both a simultaneous fit over all four $\tau^+$ decay modes and individual fits for each mode. The simultaneous fit yields a signal count of $94 \pm 31$ events, corresponding to a branching fraction of $(1.24 \pm 0.41)\times10^{-4}$, where the uncertainty is still only statistics.

\section{Systematic Uncertainties}
\label{sec:sys}

The main sources of systematic uncertainty affecting the measurement are described in the following section.\\
The largest contribution comes from limited simulation statistics, which result in a 13.3\% uncertainty. This is estimated by fluctuating the bin contents of the two-dimensional histogram PDFs 200 times according to Poisson statistics and evaluating the spread of the resulting fitted branching fractions.\\
Corrections applied to the fit variable PDFs, particularly the \ng multiplicity distributions, introduce an additional 5.5\% uncertainty. This is assessed by generating 100 sets of correction factors, each with Gaussian fluctuations derived from control sample uncertainties, and performing fits using the corresponding PDFs.\\
Other sources of systematic uncertainty include uncertainties in the branching fractions of decay modes used in the Monte Carlo (4.1\%), the efficiency of the tag-side \Bub reconstruction (2.2\%), the continuum reweighting procedure (1.9\%), \piz reconstruction efficiency (0.9\%), continuum normalization (0.7\%), and particle identification (0.6\%).\\
The uncertainties that only affect the branching fraction calculation, and not the signal yield, are estimated separately. These are the number of \FourS mesons (1.5\%), the fraction of \BpBm pairs (2.1\%), and the tracking efficiency (0.2\%).\\
Adding all contributions in quadrature, the total systematic uncertainty on the branching fraction amounts to 15.5\%.\\
The signal significance is evaluated by convolving the likelihood profile with a Gaussian distribution of width equal to the total systematic uncertainty. The test statistic is defined as $-2\log(\mathcal{L}/\mathcal{L}_0)$, where $\mathcal{L}$ is the likelihood with the signal yield floated and $\mathcal{L}_0$ is the likelihood with the signal yield fixed to zero. To compute the $p$-value, $10^6$ pseudo-experiments are generated from the background-only hypothesis, and the fraction of fits yielding a test statistic smaller than that observed in data is used. This procedure results in a signal significance of $3.0\sigma$.
\section{Conclusions}

\begin{figure}[htbp]
    \centering
    \includegraphics[width=0.5\linewidth]{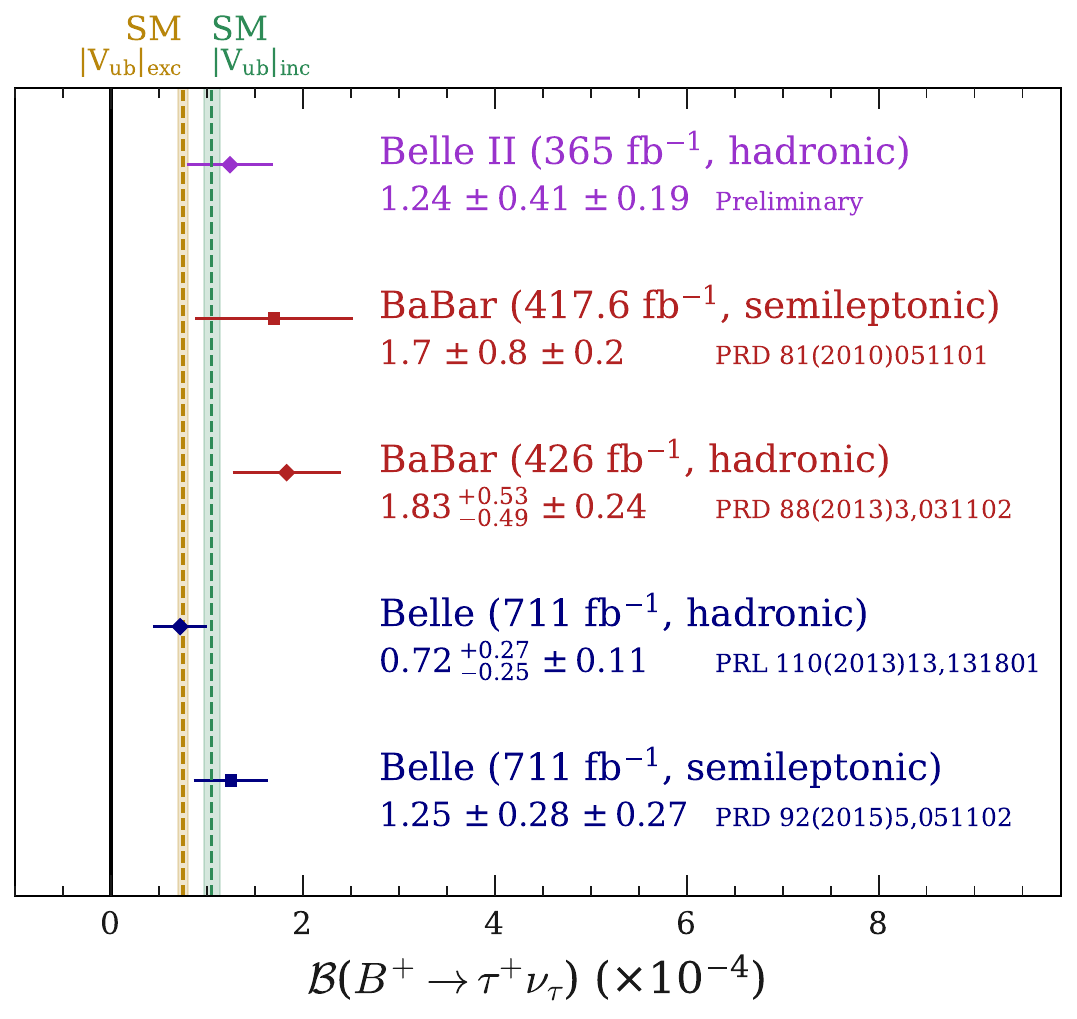}
    \caption{Branching fraction \brbtaunu measured by Belle~II compared with the past measurements and the two SM expectation values, the yellow band calculated using the exclusive value $\Vub =(3.75\pm0.06\pm0.19)\times10^{-3}$ and the green band with the inclusive value $\Vub =(4.06\pm0.12\pm0.11)\times10^{-3}$.}
    \label{fig:br_comparison}
\end{figure}

We presented a measurement of the branching fraction of the \Btaunu decay using $\SI{365}{\invfb}$ of $e^+e^-$ collision data collected at the \FourS resonance by the \belletwo detector, employing a hadronic $B$-tagging technique. Only one-prong decays of the \taup lepton are considered in this analysis.\\
We measure:
\begin{equation}
   \brbtaunu = [1.24 \pm 0.41\stat \pm 0.19\syst] \times 10^{-4},
\end{equation}
Figure~\ref{fig:br_comparison} shows a comparison of our \brbtaunu measurement with past measurements from \babar and \belle, and SM predictions based on exclusive and inclusive determinations of \Vub~\cite{banerjee2024averagesbhadronchadrontaulepton}.\\
\\
Assuming the Standard Model, and using $f_B = \SI{190.0 \pm 1.3}{\mev}$~\cite{aoki2024flagreview2024}, we extract the CKM matrix element:
\begin{equation}
    \Vub_{\Btaunu} = [4.41^{+0.74}_{-0.89}] \times 10^{-3}.
\end{equation}
Although based on a smaller data set, the statistical uncertainty of this result is comparable to previous hadronic tag measurements from \babar ($\SI{426}{\invfb}$)~\cite{babar_had} and \belle ($\SI{711}{\invfb}$)~\cite{belle_had}, thanks to improved tagging algorithms and optimized selection criteria.


\section*{References}

\begin{thebibliography}{99}

\bibitem{Belle-II:2025ruy}
I.~Adachi \textit{et al.} [Belle II Collaboration],
``Measurement of $B^+\to\tau^+\nu_\tau$ branching fraction with a hadronic tagging method at Belle II'',
\href{https://arxiv.org/abs/2502.04885}{arXiv:2502.04885 [hep-ex]}.

\bibitem{2024pdg}
S.~Navas \textit{et al.} [Particle Data Group],
``Review of Particle Physics'',
Phys. Rev. D \textbf{110}, no.3, 030001 (2024),
\href{https://doi.org/10.1103/PhysRevD.110.030001}{doi:10.1103/PhysRevD.110.030001}.

\bibitem{aoki2024flagreview2024}
Y.~Aoki \textit{et al.},
``FLAG Review 2024'',
\href{https://arxiv.org/abs/2411.04268}{arXiv:2411.04268 [hep-lat]}.

\bibitem{banerjee2024averagesbhadronchadrontaulepton}
Sw.~Banerjee \textit{et al.},
``Averages of $b$-hadron, $c$-hadron, and $\tau$-lepton properties as of 2023,''
\href{https://arxiv.org/abs/2411.18639}{arXiv:2411.18639 [hep-ex]}.

\bibitem{Jung_2010}
M.~Jung, A.~Pich and P.~Tuzón,
``Charged-Higgs phenomenology in the aligned two-Higgs-doublet model'',
JHEP \textbf{11}, 003 (2010),
\href{https://doi.org/10.1007/JHEP11(2010)003}{doi:10.1007/JHEP11(2010)003}.

\bibitem{Crivellin:2012ye}
A.~Crivellin, C.~Greub and A.~Kokulu,
``Explaining $B \to D \tau \nu$, $B \to D^* \tau \nu$ and $B \to \tau \nu$ in a 2HDM of type III'',
Phys. Rev. D \textbf{86}, 054014 (2012),
\href{https://doi.org/10.1103/PhysRevD.86.054014}{doi:10.1103/PhysRevD.86.054014}.

\bibitem{Haller:2018nnx}
J.~Haller \textit{et al.},
``Update of the global electroweak fit and constraints on two-Higgs-doublet models'',
Eur. Phys. J. C \textbf{78}, no.8, 675 (2018),
\href{https://doi.org/10.1140/epjc/s10052-018-6131-3}{doi:10.1140/epjc/s10052-018-6131-3}.

\bibitem{babar_had}
J.~P.~Lees \textit{et al.} [\babar Collaboration],
``Evidence of $B^+ \rightarrow \tau^+ \nu$ decays with hadronic B tags'',
Phys. Rev. D \textbf{88}, 031102 (2013),
\href{https://doi.org/10.1103/PhysRevD.88.031102}{doi:10.1103/PhysRevD.88.031102}.

\bibitem{belle_had}
K.~Hara \textit{et al.} [\belle Collaboration],
``Evidence for $B^- \rightarrow \tau^- \bar{\nu}_\tau$ with a Hadronic Tagging Method Using the Full Data Sample of Belle'',
Phys. Rev. Lett. \textbf{110}, 131801 (2013),
\href{https://doi.org/10.1103/PhysRevLett.110.131801}{doi:10.1103/PhysRevLett.110.131801}.

\bibitem{belle_semi}
B.~Kronenbitter \textit{et al.} [\belle Collaboration],
``Measurement of the branching fraction of $B^+ \rightarrow \tau^+ \nu_\tau$ decays with the semileptonic tagging method'',
Phys. Rev. D \textbf{92}, 051102 (2015),
\href{https://doi.org/10.1103/PhysRevD.92.051102}{doi:10.1103/PhysRevD.92.051102}.

\bibitem{babar_semi}
B.~Aubert \textit{et al.} [\babar Collaboration],
``Search for $B^+ \rightarrow \ell^+ \nu_\ell$ recoiling against $B^- \rightarrow D^0 \ell^- \bar{\nu} X$'',
Phys. Rev. D \textbf{81}, 051101 (2010),
\href{https://doi.org/10.1103/PhysRevD.81.051101}{doi:10.1103/PhysRevD.81.051101}.

\bibitem{Belle-II:2024vuc}
I.~Adachi \textit{et al.} [\belletwo Collaboration],
``Measurement of the integrated luminosity of data samples collected during 2019--2022 by the Belle II experiment'',
Chin. Phys. C \textbf{49}, 013001 (2025),
\href{https://doi.org/10.1088/1674-1137/ad806c}{doi:10.1088/1674-1137/ad806c}.

\bibitem{Kuhr:2018lps}
T.~Kuhr \textit{et al.} [\belletwo Framework Software Group],
``The Belle II Core Software'',
Comput. Softw. Big Sci. \textbf{3}, no.1, 1 (2019),
\href{https://doi.org/10.1007/s41781-018-0017-9}{doi:10.1007/s41781-018-0017-9}.

\bibitem{belleiicollaboration2020calibration}
F.~Abudinén \textit{et al.} [Belle II Collaboration],
``A calibration of the Belle II hadronic tag-side reconstruction algorithm with $B \rightarrow X\ell \nu$ decays'',
\href{https://arxiv.org/abs/2008.06096}{arXiv:2008.06096 [hep-ex]}.

\bibitem{Svidras:473367}
H.~Svidras \textit{et al.},
``Measurement of the data to MC ratio of photon reconstruction efficiency of the Belle II calorimeter using radiative muon pair events'',
BELLE2-NOTE-PL-2021-008, \href{https://doi.org/10.3204/PUBDB-2021-05692}{doi:10.3204/PUBDB-2021-05692}.


%
%
%
%
%
%
%
%
%
%

\end{thebibliography}
\end{document}